\documentclass[prb,preprint,graphicx,showpacs,preprintnumbers,amsmath,amssymb]{revtex4-1}

\usepackage{graphicx}
\usepackage{subfigure}
\usepackage{amsmath}
\usepackage{multirow}

\usepackage{dcolumn}
\usepackage{bm}
\usepackage{algorithm}
\usepackage{algpseudocode}

\newif\ifboldnumber
\newcommand{\boldnext}{\global\boldnumbertrue}

\algrenewcommand\alglinenumber[1]{%
  \footnotesize\ifboldnumber\bfseries\fi\global\boldnumberfalse#1:}

\draft 

\begin{document}


\title{SGO: An ultrafast engine for ab initio atomic structure global optimization by differential evolution} 


\author{Zhanghui Chen}
\email{zhanghuichen88@gmail.com}
\affiliation{Materials Sciences Division, Lawrence Berkeley National Laboratory, One Cyclotron Road, Mail Stop 50F, Berkeley, California 94720, United States}
\affiliation{State Key Laboratory of Superlattices and Microstructures, Institute of Semiconductors, Chinese Academy of Sciences, P.O. Box 912, Beijing 100083, People's Republic of China}
\author{Weile Jia}
\affiliation{Materials Sciences Division, Lawrence Berkeley National Laboratory, One Cyclotron Road, Mail Stop 50F, Berkeley, California 94720, United States}
\author{Lin-Wang Wang}
\email{lwwang@lbl.gov}
\affiliation{Materials Sciences Division, Lawrence Berkeley National Laboratory, One Cyclotron Road, Mail Stop 50F, Berkeley, California 94720, United States}

\date{\today}

\begin{abstract}
 This paper presents a fast method for global search of atomic structures at ab initio level. The structures global optimization (SGO) engine consists of a high-efficiency differential evolution algorithm, accelerated local relaxation methods and an ultrafast plane-wave density functional theory code run on GPU machines. It can search the global-minimum configurations of crystals, two-dimensional materials and quantum clusters without symmetry restriction in a very short time (half or several hours). The engine is also able to search the energy landscape of a given system, which is useful for exploration of materials properties for emerging applications. The exploration of carbon monolayer and platinum atomic clusters found several new stable configurations.
\end{abstract}


\maketitle

\section{Introductions}

Global search of atomic structures \cite{catlow1,uspex1,Wales2001,LBv,materialsdesign,sphere} is very useful in material science, especially for the exploration of new materials that play important roles in emergent applications such as energy-storage \cite{energy1}, superconductivity\cite{superconductivity,superconductor2}, photocatalytic water splitting \cite{water1,water2} and high-pressure materials \cite{superhard1,pressure1,pso2,pso3}. Due to the complex energy landscape, the theoretical prediction of atomic structures is extremely difficult \cite{catlow1,RMP2000}. The conventional global search methods need tens of thousands of atomic structure local relaxations to locate the target minimum \cite{catlow1,uspex1,BH1}. Since each local relaxation is performed by density functional theory (DFT) calculations or other ab initio methods, the whole search procedure is very time consuming \cite{NatPhy2009,RMP2000}. A simple search job could cost several days or even several weeks. Therefore, it is critical to accelerate the search by highly efficient algorithms.

There have been some groups that developed efficient global search engine for atomic structures. For example, Oganov et al. developed a package USPEX (Universal Structure Predictor: Evolutionary Xtallography) for crystal structure prediction based on genetic algorithm \cite{uspex1,uspex2}. Ma et al. set up a crystal searching method through particle-swarm optimization, and obtained great success in high-pressure materials \cite{pso1,pso2,pso3}. Catlow et al. used genetic algorithm to generate plausible crystal structures from the knowledge of only the unit cell dimensions and constituent elements \cite{catlow1,catlow2}. Zunger and Zhang et al. used genetic algorithm and global space-group optimization for inverse design of materials \cite{zunger1,zunger2}. Besides, basin/minima hopping algorithms have been adopted by many groups for atomic clusters¡¯ searches \cite{BH1,BH2,BH3}. Our own PDECO (parallel differential evolution cluster optimization) procedure also showed superior performance in the search of various clusters \cite{pdeco}. These activities in this area indicate the importance of high-efficiency search algorithms for crystals, clusters and other atomic structures from emergent applications.

In this paper, we present an ultrafast engine for structures global optimization (SGO) \cite{sgo} with acceleration on all the levels of searches. The algorithm on the level of global search uses parallel differential evolution (DE) that has been demonstrated to be of high efficiency in searching structures \cite{pdeco,DE1,DE2}. On the middle level, the local relaxation of atomic structures employ curved-line-search (CLS) method \cite{CLS} and preconditioned conjugate gradient (PCG) algorithm \cite{PCG}, both of which are based on force fitting and have been demonstrated to be 2-6 times faster than conventional conjugate gradient (CG) algorithm \cite{CG,vasp}. On the level of the energy and force evaluation, we use an ultrafast DFT plane-wave code run on GPU machines \cite{pwmat1,pwmat2,GPU1,GPU2}. The integration of three-level acceleration enables the ultrafast global search of atomic structures. Symmetry restriction that is widely used for acceleration in other packages \cite{uspex1,pso1,pso2} is not required here. Tests for crystals, two-dimensional (2D) materials and quantum clusters demonstrate the superior search performance of SGO engine. Moreover, this engine can also search the energy landscape of given systems, which is useful for exploration of materials properties.

The paper is organized as follows. Section 2 introduces the search engine with the three-level acceleration algorithms. Section 3 presents the tests on crystals, 2D materials and quantum clusters as well as the searches of energy landscape. The concluded remarks are drawn in Section 4.

\section{Methods}

\subsection{Overview of the SGO engine}

\begin{figure} [!t]
\centering
\includegraphics[scale=0.83]{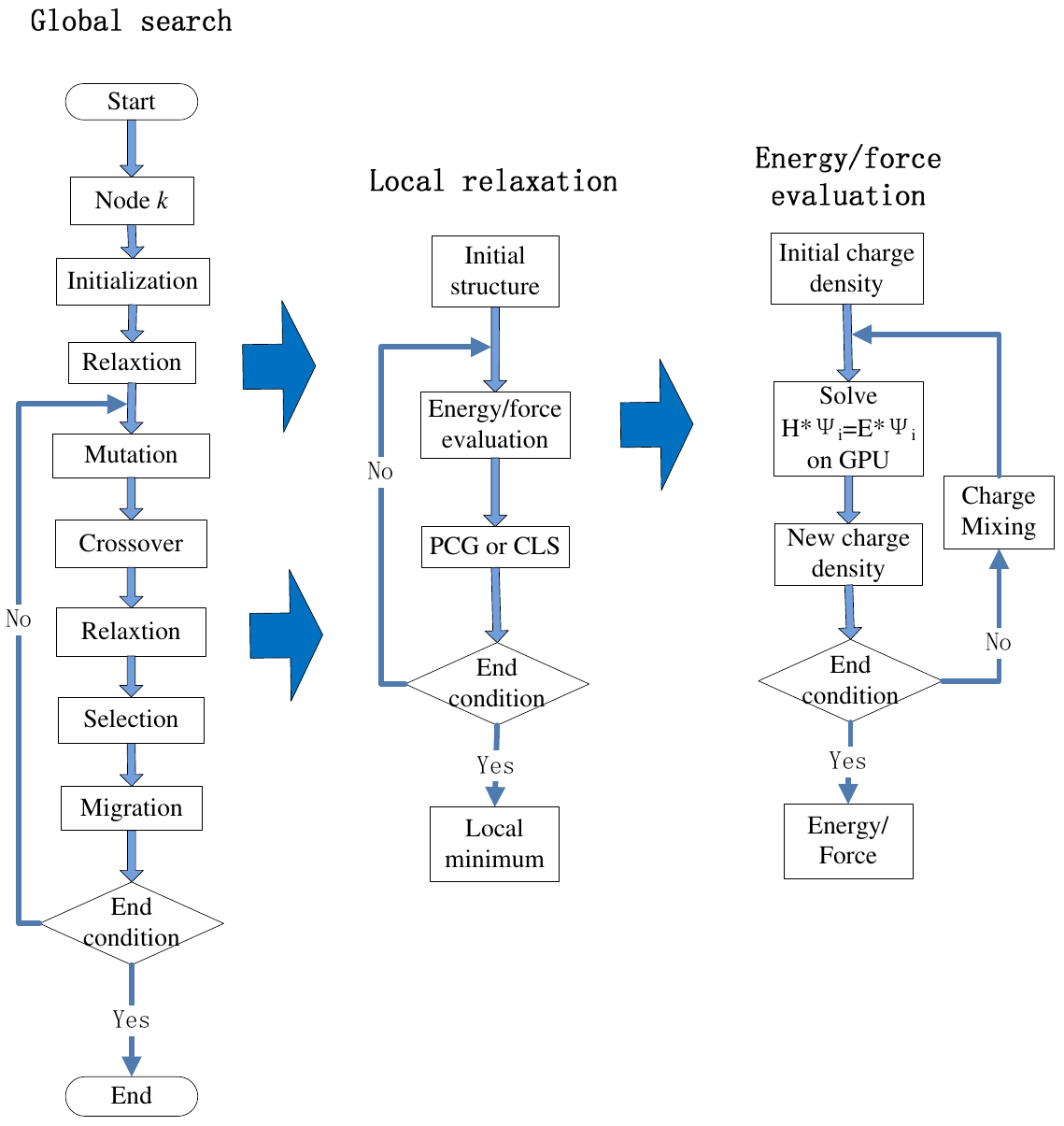}
\caption{Flow chart of the SGO engine. It consists of three modules: global search (left), local relaxation (middle) and energy/force evaluation (right).}\label{slab}
\end{figure}

Our search engine SGO consists of three levels, as shown in Figure 1. On the top level, we employ an improved parallel DE algorithm for global search. On the middle level, CLS and PCG algorithms are employed for accelerated local relaxation of each atomic structure into its nearest local minimum. On the bottom level, the energy and atomic force of each structure are calculated self-consistently by the DFT plane-wave code run on GPU machines.

\subsection{Fast global search}

The global search is implemented by a high-efficiency parallel DE algorithm. DE \cite{DE1,DE2,DE3,DE4,DE5,DE6,DE7} belongs to the family of evolutionary algorithms (EAs)\cite{EA1,EA2,EA3}. It employs multiple individuals' cooperation and evolutionary operators to find the global minimum of the objective function. Compared to most other EAs, it is much more simple and straightforward to implement. The main search engine requires less than 30 lines of C or FORTRAN code \cite{DE2,DE7}. Due to very robust control parameters and low space complexity \cite{DE2,DE4,DE7}, DE is highly efficient. Since its proposal in 1995 \cite{DE1}, DE continues to secure front ranks in various competitions and exhibits superior performance than all the other EAs in most of optimization problems \cite{DE1,DE2,DE3,DE4,DE5,DE6,DE7}.

Assuming an optimization problem is formulated as: $\mathop {\min }\limits_X f\left( X \right)$, where $X$ is a vector of $1 \times n$ variables, DE works with a population of $N_p$ candidate solutions, i.e., $X_{i,G}$, $i = 1, ... ,N_p$, where $i$ is the index of the individual and $G$ is the generation in the evolution. Random initialization, mutation, crossover and selection operators are then employed to evolve the population. These operators are similar as other EAs except the mutation one which applies the vector differential between the existing population members for determining both the degree and direction of the perturbation to the individual subject. A conventional \emph{differential mutation} is designed to be a linear combination between a randomly selected \emph{base individual} and a scaled difference between two other \emph{donor individuals}:
\begin{equation}
{V_{i,G}} = {X_{r3,G}} + F\left( {{X_{r1,G}} - {X_{r2,G}}} \right)
\end{equation}
where $i = 1, \cdots , N_p$; $r1$, $r2$, $r3$ $\in {1, \cdots ,N_p}$ are randomly selected and satisfy $r{\rm{1}} \ne r{\rm{2}} \ne r{\rm{3}} \ne i$; ${V_{i,G}} = \left( {{v_{1,i,G}}, \cdots ,{v_{j,i,G}}, \cdots ,{v_{n,i,G}}} \right)$ is the perturbed individual and the scaling factor $F$ ($F \in [0,{\rm{1}}]$) is a control parameter.

Our DE search engine is an improved version of the conventional procedure based on the characteristics of atomic structures optimization. The complete search flow chart (left module of Figure 1) consists of initialization of individuals (i.e., atomic structures) on each node followed by local relaxation to their nearest local minimum. Then, improved mutation and crossover operators are implemented to produce possible descendant which are further locally relaxed. We then use greedy selection and migration operators to determine the final descendant to next evolution. Such procedure is repeated until the global minimum is found or end condition is reached.

In details, the population initialization includes the following operations. Firstly, the supercell box size of each atomic structure is generated by a slightly random perturbation of the values set by the user. The atoms are then randomly arranged inner the supercell. The atom coordinates are constricted depending on the type of systems (atomic clusters, 2D materials, 3D periodic systems, adsorption systems, etc.). For example, atoms will be placed at a slab space for 2D materials. To guarantee the individual quality and population diversity, SGO will check distance between atoms in one structure and check similarity between different structures. Chaos operator and growing operator (see Ref.~\citenum{pdeco} for details) are also considered to improve the sampling.

For the mutation scheme, the following triangle differential mutation \cite{DE3} is used:
\begin{equation}
\begin{array}{l}
{V_{i,G + 1}} = {{\left( {{X_{r1,G}} + {X_{r2,G}} + {X_{r3,G}}} \right)} \mathord{\left/
 {\vphantom {{\left( {{X_{r1,G}} + {X_{r2,G}} + {X_{r3,G}}} \right)} 3}} \right.
 \kern-\nulldelimiterspace} 3} + \left( {{p_2} - {p_1}} \right)\left( {{X_{r1,G}} - {X_{r2,G}}} \right)\\
{\kern 1pt} {\kern 1pt} {\kern 1pt} {\kern 1pt} {\kern 1pt} {\kern 1pt} {\kern 1pt} {\kern 1pt} {\kern 1pt} {\kern 1pt} {\kern 1pt} {\kern 1pt} {\kern 1pt} {\kern 1pt} {\kern 1pt} {\kern 1pt} {\kern 1pt} {\kern 1pt} {\kern 1pt} {\kern 1pt} {\kern 1pt} {\kern 1pt} {\kern 1pt} {\kern 1pt} {\kern 1pt} {\kern 1pt} {\kern 1pt}  + \left( {{p_3} - {p_2}} \right)\left( {{X_{r2,G}} - {X_{r3,G}}} \right) + \left( {{p_1} - {p_3}} \right)\left( {{X_{r3,G}} - {X_{r1,G}}} \right)
\end{array}
\end{equation}
where ${p_1} = {{\left| {f({X_{r1,G}})} \right|} \mathord{\left/
 {\vphantom {{\left| {f({X_{r1,G}})} \right|} {p'}}} \right.
 \kern-\nulldelimiterspace} {p'}}$, ${p_2} = {{\left| {f({X_{r2,G}})} \right|} \mathord{\left/
 {\vphantom {{\left| {f({X_{r2,G}})} \right|} {p'}}} \right.
 \kern-\nulldelimiterspace} {p'}}$, ${p_3} = {{\left| {f({X_{r3,G}})} \right|} \mathord{\left/
 {\vphantom {{\left| {f({X_{r3,G}})} \right|} {p'}}} \right.
 \kern-\nulldelimiterspace} {p'}}$ and  $p' = \left| {f({X_{r1,G}})} \right|{\kern 1pt} {\kern 1pt} {\kern 1pt}  + {\kern 1pt} {\kern 1pt} \left| {f({X_{r2,G}})} \right|$ $+ \left| {f({X_{r3,G}})} \right|$. $i = 1, \cdots , N_p$; $r1$, $r2$, $r3$ $\in {1, \cdots ,N_p}$ are randomly selected and satisfy $r{\rm{1}} \ne r{\rm{2}} \ne r{\rm{3}} \ne i$. Compared to Formula (1), this scheme is shown to well maintain the balance between the local convergent speed and the global search possibility when the population size is small.

For the crossover scheme, the conventional multi-point crossover is replaced by the ¡°cut-and-splice¡± crossover \cite{sphere,crossover} which can makes use of the three-dimensional space distribution information of the parent structures. SGO provides two schemes: plane-cut-splice and sphere-cut-splice. Assuming the two parent structures are noted as P1 and P2, the first scheme chooses a random plane passing through the mass center of each structure. Then it cuts the structures on this plane, and assembles one child from the atoms of P1 which lie above the plane, and the atoms of P2 which lie below the plane. If the child does not contain the correct number of atoms, the operator rotates the plane until the generated offspring contain the correct number of atoms. The second scheme chooses a random sphere around each structure. Parent structures exchange their atoms inner the sphere to produce the offspring. It is found that the sphere scheme has better performance in large systems or core-shell systems \cite{sphere}. Figure 2(a) and 2(b) illustrate the plane-cut-splice and sphere-cut-splice crossover, respectively.

\begin{figure} [htbp]
\centering
\includegraphics[scale=0.45]{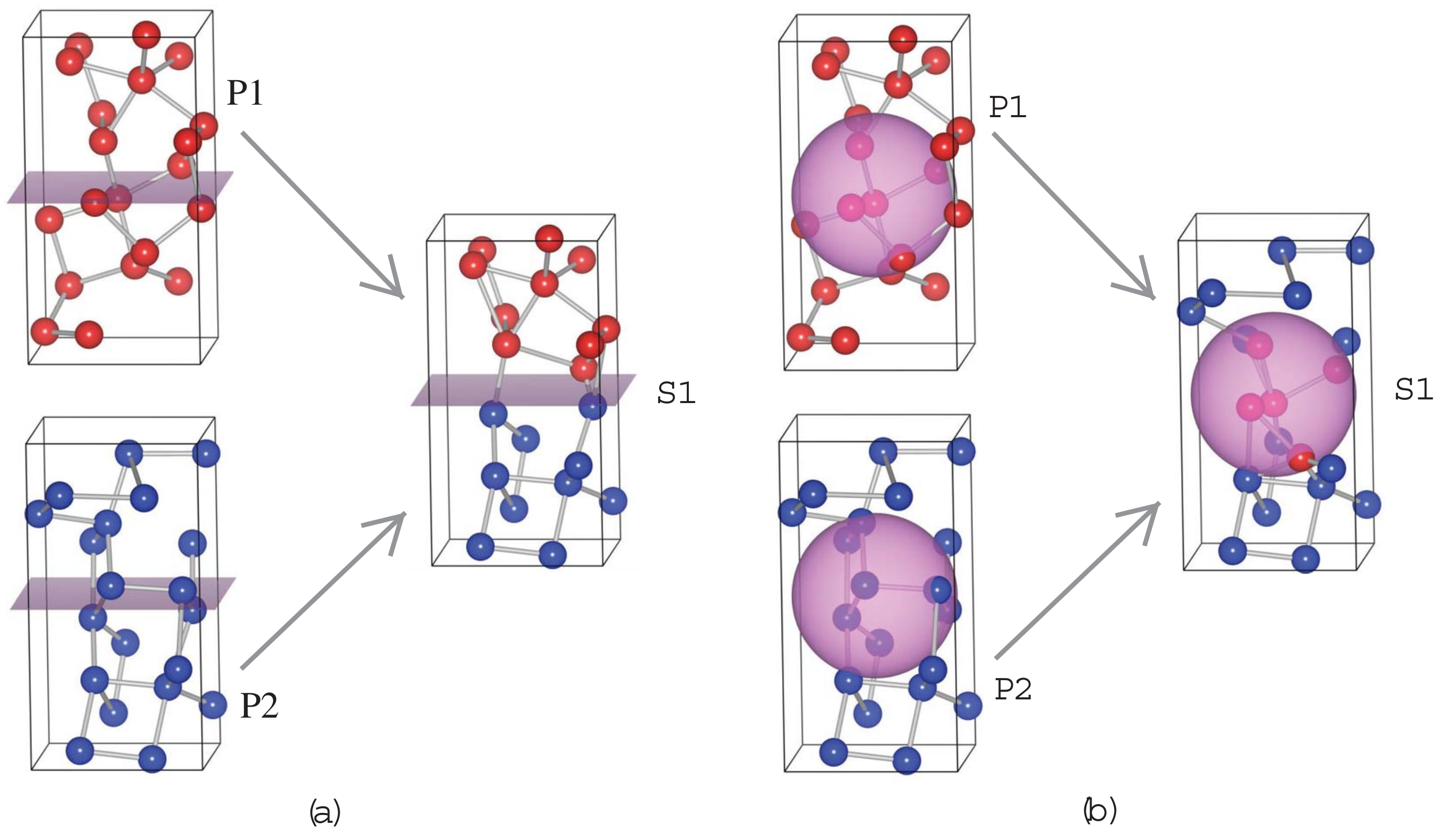}
\caption{Schematic diagram for the cut-and-splice crossover. (a) is the plane scheme and (b) is the sphere scheme. P1 and P2 are the two parent structures while S1 is the generated offspring structure.}\label{slab}
\end{figure}

For the selection scheme, SGO uses a simple greedy operator, i.e., the lower-energy one between parents and offsprings survives. For the parallel scheme, SGO employs a subpopulation-type parallelization. The whole population is partitioned into several subpopulations, each of which communicates with others by the migration operator \cite{DE6,DE7}. Such improvements have shown better search efficiency. More details about these evolutionary operators can be found in Ref.~\citenum{sphere} and \citenum{pdeco}.

\subsection{Accelerated local relaxation}

The initial structure passed to the module of local relaxation (middle module of Figure 1) is usually far from its nearest local minimum. Conventional relaxation methods, such as CG algorithm \cite{CG}, need hundreds of steps to reach convergence \cite{CLS,PCG}. It is thus crucial to speed up this process.

We have proposed two accelerated local relaxation algorithms in Ref.~\citenum{CLS} and Ref.~\citenum{PCG}. One is CLS algorithm \cite{CLS}, which is based on on-the-flight force learning and a corresponding curved line minimization algorithm. CLS algorithm firstly call a non-linear force-fitting procedure to improve the classical force field to reproduce the DFT atomic forces. The fitted force field provides a guided search curve from the initial configuration to its minimum. CLS then implements ab initio line minimization along this guided curve on DFT energy landscape. The algorithm repeats force fitting, constructing guided search curve and ab initio line minimization until locating the DFT local minimum.

The other one is PCG algorithm\cite{PCG}, in which the pre-conditioner is provided by an approximate inverse Hessian matrix. In details, PCG also firstly call force-fitting procedure to improve the parameters of classical force field, so as to reproduce the DFT atomic forces. Then, the corresponding Hessian matrix is given by calculating the second-order finite difference of the force-fitted potential energy. Because there are three zero eigenvalues for Hessian matrix corresponding to three translational modes, we constructed the inverse Hessian matrix $H^{-1}$ as $\sum\nolimits_j^{{\tau _j} \ne 0} {{{V_j^T{V_j}} \mathord{\left/
 {\vphantom {{V_j^T{V_j}} {{\tau _j}}}} \right.
 \kern-\nulldelimiterspace} {{\tau _j}}}} $ , where $V_j$ and $\tau _j$ are the eigenvector and eigenvalue of $H$, respectively. Finally, this inverse Hessian matrix is taken as the pre-conditioner of the common PCG procedure. Such force-fitted pre-conditioner might be updated during the relaxation process, depending on how far away the initial structure is from its local minimum.

Both CLS and PCG algorithms have shown a speedup factor of 2 to 6 on the relaxation of most atomic systems. CLS, PCG, conventional CG and Broyden-Fletcher-Goldfarb-Shanno (BFGS) \cite{BFGS} algorithms are provided to the user as the input relaxation option. Note that SGO does not require a high-precision local relaxation, especially at the initial stage of the evolution. Parameter such as rough K-points sampling grid, small plane-wave energy cutoff and low force convergence criteria (e.g., 0.02 eV/$\AA$) can be used to save the computational cost.

\subsection{DFT plane-wave code on GPU}

The computations of energy and atomic force of each structure (right module of Figure 1) are performed by PWmat \cite{pwmat1,pwmat2,GPU1,GPU2}, a DFT \textbf{P}lane \textbf{W}ave \textbf{mat}erials simulation code run on GPU clusters. PWmat is developed based on a CPU DFT plane-wave psedudopotentials code: PEtot \cite{PEtot}. Compared to PEtot, lots of implementations have been greatly improved in terms of algorithms¡¢ and stabilities. It has simple input, very easy to use. It supports a serial of tested set of pseudopontentials such as norm-conserving, utrasoft and PBE. The current PWmat version (v1.5) \cite{pwmat1} has $18\thicksim30$ times of speedup over the CPU PEtot code \cite{GPU1, GPU2}. 

\begin{algorithm}[H]
  \caption{ All-band conjugate-gradient (CG) method for solving Kohn-Sham equation $\hat{H}|\Psi_i\rangle=\varepsilon_i|\Psi_i\rangle$ in GPU. The time consuming steps indicated by bold numbers are implemented by CUBLAS library, wavefunction data compression or hybrid parallelization of GPU and CPU. Others are implemented by hand-written CUDA kernel subroutines.}
  \label{ABCG}
  \begin{algorithmic}[1]
    \State $k \gets 0$
    \boldnext \State Calculate $\hat{H}|\Psi_i\rangle $ and $H(i,j)=\langle \Psi_i|\hat{H}|\Psi_j\rangle$.
    \boldnext \State Calculate the eigen value $\varepsilon_i$ by $H$ matrix subspace diagonalization.
    \While{($\sim$ convergence)}
    \State $k \gets k+1$
    \boldnext \State Residual: $P_i(k)=\hat{H}|\Psi_i\rangle -\varepsilon_i|\Psi_i\rangle $
    \If {$k = 1$}
      \State Preconditioned: $P_i(k)=A\cdot P_i(k)$. $A$ is the preconditioned matrix.
    \Else
       \State Preconditioned-CG: $P_i(k)=A\cdot \left( P_i(k)-\alpha_i \cdot P_i(k-1)\right)$. $\alpha_i$ is CG coefficient.
    \EndIf
    \boldnext \State Projection: $P_i(k)=P_i(k)- \sum\nolimits_{j = 1,i} | \Psi_j \rangle \langle \Psi_j | P_i \rangle $
    \State Update wavefunction by line minimization: $|\Psi_i\rangle=|\Psi_i\rangle \cos \theta_i+P_i(k) \sin \theta_i$.
    \boldnext \State Orthogonalization: $|\Psi_i\rangle=|\Psi_i\rangle - \sum\nolimits_{j = 1,i-1} | \Psi_j \rangle \langle \Psi_j | \Psi_i \rangle$
    \boldnext \State Update $\hat{H}|\Psi_i\rangle $ and $H(i,j)=\langle \Psi_i|\hat{H}|\Psi_j\rangle$.
    \boldnext \State Update the eigen value $\varepsilon_i$ by $H$ matrix subspace diagonalization.
    \EndWhile
  \end{algorithmic}
\end{algorithm}

This is achieved mainly by adapting the self-consistent electronic structure calculation to the heterogeneous computer architecture. One self-consistent step begins with an initial charge density $\rho_{in}$, then Kohn-Sham equation is solved to get the converged wavefunction $\Psi_i$ and the output charge density $\rho_{out}$. $\rho_{out}$ is then mixed with $\rho_{in}$ for the next step. In this procedure, the most time consuming computation is to iteratively solve Kohn-Sham equation by the all-band conjugate-gradient (CG) method. This method begins with a initial set of wavefunctions, Hamiltonian matrix $\langle \Psi_i|\hat{H}|\Psi_j\rangle$ and the corresponding eigen values $\varepsilon_i$. Then the residual $P_i$ is calculated and transformed to a preconditioned CG vector, which is further projected by wavefunctions. Such CG vector is used to relax the wavefunction by a line minimization procedure. Updated wavefunction is then orthogonalized and used to update the Hamiltonian matrix. Finally new eigen values are calculated by subspace diagonalization. Such CG steps will be repeated several times until the residual $P_i$ reaches the convergence condition. Flow chart of this all-band CG method is detailed in Algorithm \ref{ABCG}. In the PWmat, we have moved the entire all-band CG method into GPU by using the CUBLAS library, hybrid parallelization, and wavefunction data compression to reduce MPI communication \cite{GPU1,GPU2}. Note that the hybrid parallelization scheme is also used in the occupation in the self-consistent charge density calculation, thus moving the entire electronic structure calculation into GPU to fully utilize the GPU computing power.

The PWmat and SGO codes can be run on common GPU machines (e.g., Titan of Oak Ridge Leadership Computing Facility) or our own custom-made GPU machines: Mstation (Material Station) \cite{pwmat1,GPU1,GPU2}. One Mstation has 4 NVIDIA Titan X GPU processors and 128 GB RAM, supporting the DFT calculations of as large as 1,000 atoms.


Note that the local relaxation of all the individuals in one generation can be implemented in parallel or in serial, depending on the available computational resources. For an example with 8-individuals population, the 8 atomic structures can relaxed at the same time by 8 chips of GPU, one of which runs one local relaxation job. This can be achieved by two Mstations. They also can be relaxed one by one serially using only 1 chip of GPU resource. In general, 8 individuals could be qualified for the search of systems with 20 atoms or less, because of the improvement of global search ability. Multiple independent runs might be needed to compensate the  stochastic feature of differential evolution algorithms.

\section{Results}

Because of acceleration on all the levels of searches, the SGO engine is expected to be very fast. We tested SGO on the search of global minimum of various systems (crystals, 2D materials and clusters), and presented its results on the search of energy landscape. The tests were run on the custom-made Mstation machine.

\subsection{Search of global minimum}

\subsubsection{Three-dimensional crystals}

We first tested SGO on some well-known crystals with one (Si and Mg), two (SiO$_2$ and ZnO) or three elements (CaCO$_3$ and BiFeO$_3$). Table 1 lists the settings and the search performance. During these global searches, the number of atoms for each element in the supercell are fixed. 

We can see that SGO can find the known ground-state structures of Si and Mg in one or two generations with a small population (8 individuals). For the other cases with two or three elements, SGO are still able to locate the targeted global minimum in less than 15 generations. These systems cost very few numbers of local relaxations and demonstrate the global search efficiency of the designed DE algorithm. Moreover, all the calculations are completed in only two hours on two Mstations.

\begin{table}[t]
\caption{\label{tab:table1}Global search of the ground-state structure of various systems. The target structure is indicated by space group for crystals and 2D materials or point group for clusters. $N$(atoms) is the number of atoms in the searching supercell. $N$(Gen) is the number of generations on average needed to locate the global minimum in multiple independent runs. $N$(LM) is the corresponding average number of local minimization. $T$(hours) is the average computational time for running on 2 Mstations. The time will be double if tests are run on only 1 Mstation. The population size is set to 8 for all the tests. }
\begin{ruledtabular}
\begin{tabular}{ccccccc}
\multicolumn{2}{c}{Systems} & Structures &	$N$(atoms) &	$N$(Gen) & $N$(LM) &$T$(hours) \\
\hline
\multirow{6}{*}{Crystals} &Si	&Fd$\bar{3}$m	&8	&1.89	& 15.1	&0.081\\
	&Mg	&P6$_3$/mmc	&8	& 1.08	& 8.64	&0.034\\
	&SiO$_2$	&P3$_1$21	&9	&5.22	&41.8	&0.37\\
	&ZnO	&F$\bar{4}$3m	&8	&3.25	&26.0	&0.36\\
	&BiFeO$_3$	&R3c	&10	&7.43	&59.4	&0.83\\
	&CaCO$_3$	&R$\bar{3}$c	&10	&12.4	&99.2	&1.53\\
\hline
\multirow{2}{*}{2D materials}	&Graphene	&P$_6$/mmm	&8	&8.75	&70.0	&0.16\\
	&MoS$_2$	&P$\bar{6}$m2	&12	&29.4	&235.2	&1.36\\
\hline
\multirow{2}{*}{Atomic clusters}	&Pt$_{10}$	&T$_d$	&10	&22.7	&181.6	&2.18\\
	&Pt$_{23}$	&C$_1$	&23	&78.0	&624.0	&10.4\\
\end{tabular}
\end{ruledtabular}
\end{table}

\begin{figure} [htbp]
\centering
\includegraphics[scale=0.8]{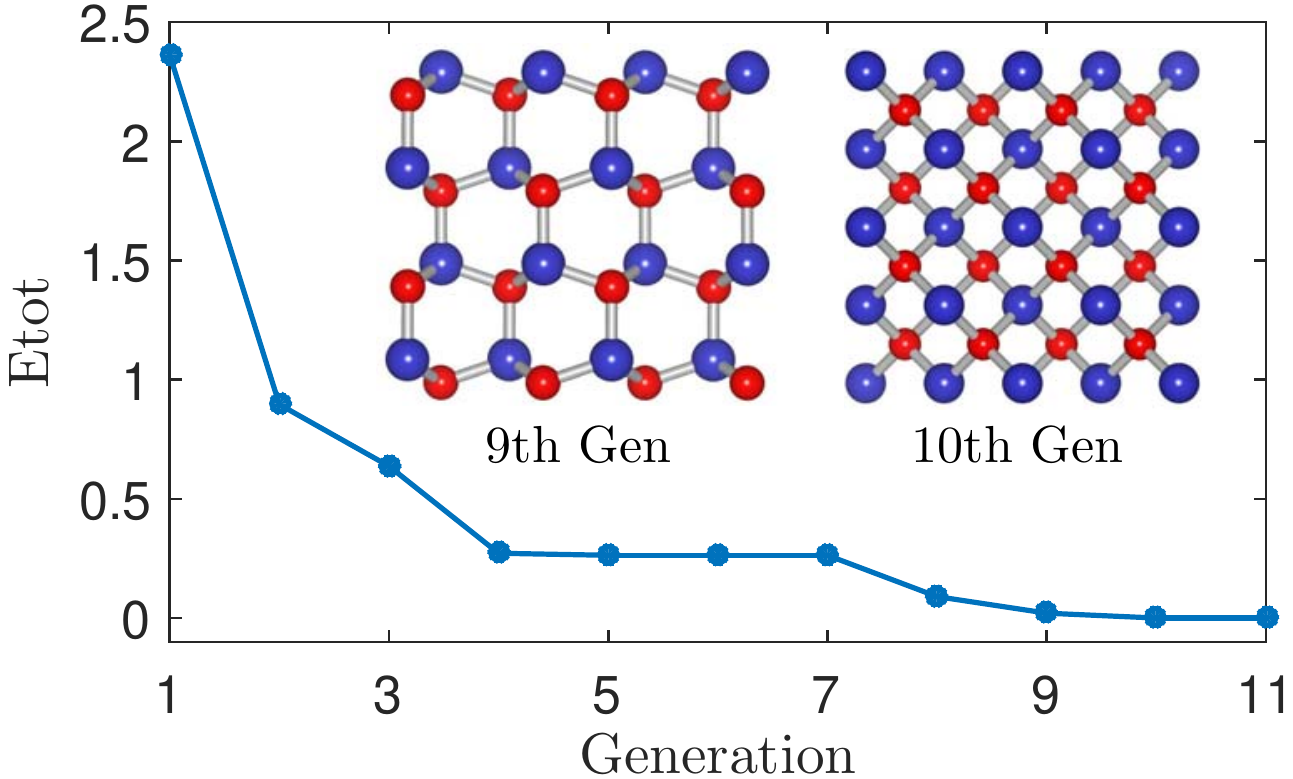}
\caption{Evolution of the best individual of the population in one search of ZnO. Etot indicates the total energy of one structure compared to the global-minimum structure in the unit of eV. SGO finds the wurtzite structure in the 9$^{th}$ generation and the targeted zincblende structure in the 10$^{th}$ generation.}
\label{fig-ZnO}
\end{figure}

Note that we do not use any symmetry restriction during the initialization and local relaxation. This type of free search is more difficult than the one using specific symmetry restriction, but is able to search all the possible configurations in the whole energy landscape. Figure \ref{fig-ZnO} shows the evolution of the best ZnO individual in the population. SGO not only finds the targeted zincblende structure (space group F$\bar{4}$3m) but also the wurtzite structure (space group P6$_3$mc). In the used PBE potential, zincblende structure has slightly lower energy than wurtzite structure.

\subsubsection{Two-dimensional materials}

Now we tested two known 2D materials: graphene and MoS$_2$. The known ground state of 2D carbon materials is a monolayer of six-member rings (i.e., graphene) while the ground state of 2D MoS$_2$ is a three-layer hexagonal structure with two S layers at both sides of Mo layer. SGO spends 8.75 and 29.4 generations to locate these two configurations, respectively. The costing time is a little longer than the one for crystals with the same number of elements because there is one free dimension without periodic restriction. SGO does not constrain atoms on a fixed plane, but allows them moving freely in a big enough slab. This free dimension will produce a larger number of possible configurations and a corresponding harder search.

\begin{figure} [htbp]
\centering
\includegraphics[scale=0.8]{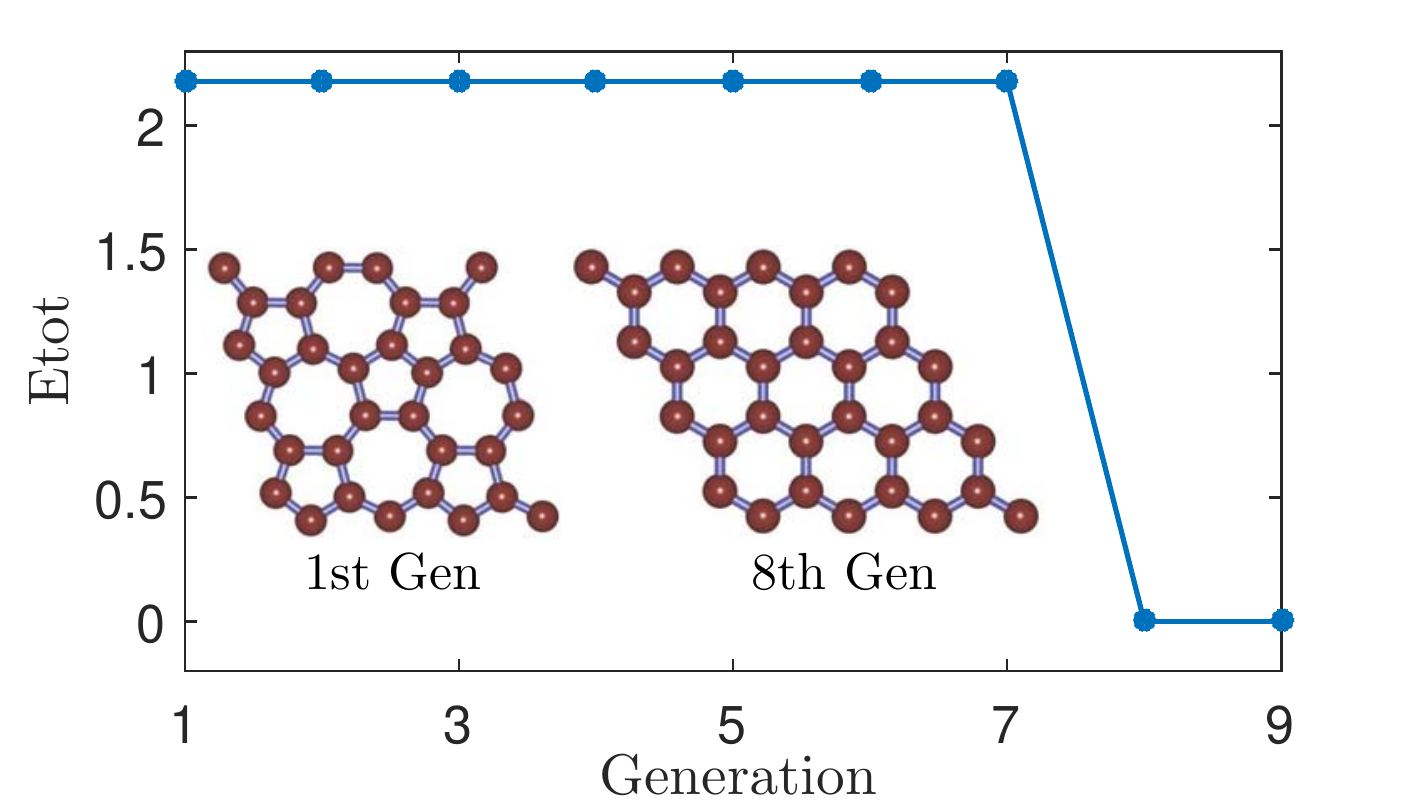}
\caption{Evolution of the best individual of the population in one search of graphene. Etot indicates the total energy of one structure compared to the global-minimum structure in the unit of eV. The configuration found in the 1$^{st}$ generation is a monolayer consisting of five-member rings and seven-member rings while the ground state found in the 8$^{th}$ generation is six-member rings' structure.}
\label{fig-graphene}
\end{figure}

Figure \ref{fig-graphene} shows the evolution of the best individual of the population in one search of graphene. This search costs 8 generations to reach the global minimum. The lowest-energy configuration in the first generation is a monolayer consisting of five-member rings and seven-member rings. This new configuration also owns to the search without symmetry restriction. It costs 7 more generations for SGO to evolve to the targeted monolayer consisting of six-member rings.

\subsubsection{Atomic clusters}

\begin{figure} [htbp]
\centering
\includegraphics[scale=0.8]{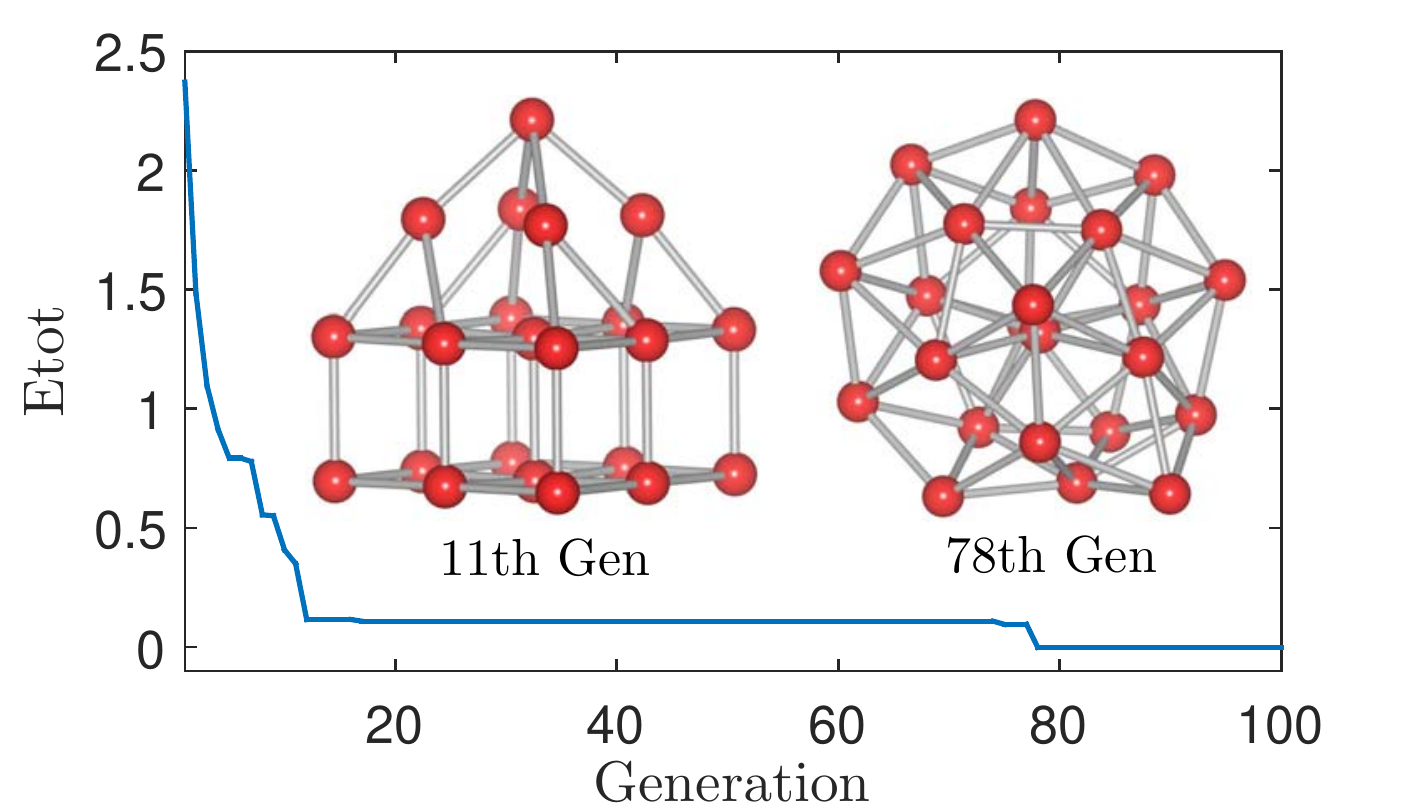}
\caption{Evolution of the best individual of the population in one search of Pt$_{23}$ clusters. Etot indicates the total energy of one structure compared to the global-minimum structure in the unit of eV. The configuration found in the 11$^{th}$ generation is a known structure with C$_{4v}$ point symmetry and the one in the 78$^{th}$ generation is the new minimum.}
\label{fig-Pt23}
\end{figure}

The energy landscape of atomic clusters is more complex than the one of crystals and 2D materials because there is no periodic boundary restriction. There will be a large number of free clusters with stable configurations. We tested SGO on two Pt clusters: Pt$_{10}$ and Pt$_{23}$. The known global minima of Pt$_{10}$ and Pt$_{23}$ have the point symmetry of T$_d$ and C$_{4v}$, respectively \cite{Wang2009}. In our searches, SGO find the T$_d$-symmetry Pt$_{10}$ in 37$^{th}$ generations and the C$_{4v}$-symmetry Pt$_{23}$ in 78$^{th}$ generations. For Pt$_{23}$, SGO even finds a few structures with lower energy. Figure \ref{fig-Pt23} shows the corresponding evolution process of Pt$_{23}$ clusters. We can see that the final minimum is about 0.35 eV lower in energy than the known C$_{4v}$ structure. This new minimum configuration has no symmetry operation.

Compared to the searches of other systems, Pt clusters cost more computer time. There are several reasons. First, the big vacuum around clusters and the large number of valance electrons in pseudopotential magnify each electronic self-consistent calculation time. Second, the complex energy landscape results in that each local relaxation needs more iterations to reach convergence and the global search needs more evolving generations. Nevertheless, several hours are still acceptable in exploring new stable configurations.

\subsection{Search of energy landscape}

\begin{figure} [!b]
\centering
\includegraphics[scale=0.8]{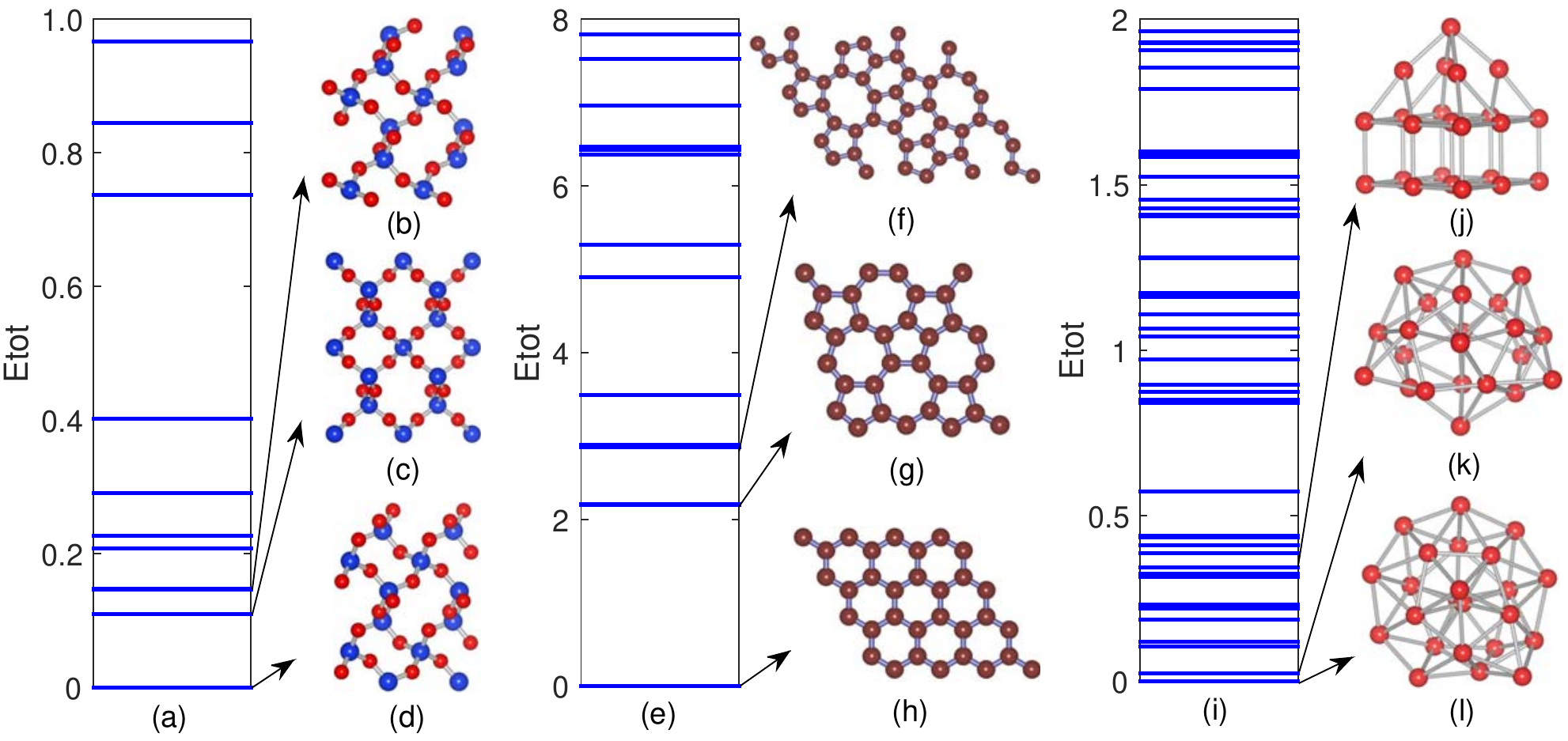}
\caption{The energy landscape of (a-d) SiO$_2$ crystal, (e-h) 2D carbon monolayer and (i-l) Pt$_{23}$ atomic cluster. Each blue line in (a), (e) and (i) indicates one configuration. (b-d), (f-g) and (j-l) are several representative configurations. Etot is the total energy of one configuration compared to the global-minimum configuration in the unit of eV.}
\label{fig-landscape}
\end{figure}

Due to the ultrafast search speed, SGO is able to explore the energy landscape of a given system in an affordable time, especially exploring the landscape around the global minimum. We tested this function on the systems of SiO$_2$ crystal, carbon monolayer and Pt$_{23}$ atomic clusters. We adopted 9-atoms supercell for SiO$_2$ and 8-atoms supercell for carbon. Pt$_{23}$ cluster is contained in a 24\AA$\times$24\AA$\times$24\AA $ $ supercell. About 500 local minimizations are searched for SiO$_2$ and carbon systems, and about 3000 for Pt clusters. Figure \ref{fig-landscape} shows their found landscape, depicted by local-minima energy spectrum.

For SiO$_2$, SGO finds a few stable crystals in the energy region of [0, 1.0] eV close to the ground state. The space group of the ground-state structure is P3$_1$21 (Figure \ref{fig-landscape}d). The configuration 0.11 eV above the ground state has a higher symmetry with space group P6$_2$22 (Figure \ref{fig-landscape}c) while the one with 0.15 eV energy has space group P3$_2$21 (Figure \ref{fig-landscape}b), very similar to the ground-state structure. Both P6$_2$22 and P3$_2$21 structures have been confirmed in other studies \cite{SiO21,SiO22}. Note that the relative energy between different configurations could be changed at different pseudopential, external pressure or temperature.

For 2D carbon monolayer, there is no other stable configuration except the well-known graphene (Figure \ref{fig-landscape}h) in the energy region of [0 2.0] eV. This indicates to some extent that there is a large energy barrier from graphene to other structures. Thus, graphene is very stable, even at high temperature or strong external interaction. The configuration (Figure \ref{fig-landscape}g) with 2.2 eV above the graphene consists of five-member and seven-member carbon rings. It has the symmetry Cmmm and is also very stable. There are two stable configurations with about 0.7 eV higher than the five-seven membered rings' structure. The lower-energy one (Figure \ref{fig-landscape}f, space group Pmmm) is made up of five-member, six-member and eight-member rings, while the higher-energy one (space group Cmmm) is made up of four-member, six-member and eight-member rings. Their energy difference is only 0.036 eV. Above these configurations, there are a few other stable isomers with different rings¡¯ structures, such as four-five-ten membered rings.

For Pt$_{23}$ cluster, the energy spectrum is much dense than the ones of SiO$_2$ crystal and carbon monolayer. There are a lot of local minima configurations in the energy region of [0 2.0] eV. Especially, a few of them have lower energy than the putative global minimum with C$_{4v}$ symmetry in literatures \cite{Wang2009} (Figure \ref{fig-landscape}(j)). Most of these lower-energy configurations have lower-symmetry geometry.
This dense local minima spectrum indicates a complex ``glass-like'' or ``fluxional'' energy landscape. Such landscape feature can be further identified by calculating energy barrier between different local minima.

Note that the energy landscapes of Figure \ref{fig-landscape} are given by a limited number of searches with a fixed number of atoms in the supercell. There could be more configurations when we use a different number of atoms and spend more searches. Such energy landscape is helpful to understand the known materials¡¯ properties and to explore new configurations.

\section{Conclusions}

An ultrafast engine SGO for atomic structure global optimization is proposed. We speeded up the engine by a high-efficiency DE algorithm, accelerated local relaxation methods and an ultrafast DFT code run on GPU machines. We tested SGO on well-known crystals, 2D materials and atomic clusters. SGO can find the targeted minimum in half or several hours. SGO are also able to explore the energy landscape of a given system in an affordable computer time. The code will be presented soon at the websites: www.sgo.ac.cn and www.pwmat.com.

\bibliography{SGO}

\end{document}